\documentclass[aps,prd,preprint,preprintnumbers,superscriptaddress,floatfix,nofootinbib]{revtex4}
\usepackage{epsfig}
\usepackage{bm}
\usepackage{amssymb}
\usepackage{amsmath}
\usepackage{color}
\usepackage{subfigure}

\begin{document}

\title{Perfect fluidity of a dissipative system: Analytical solution for the
Boltzmann equation in $\mathrm{AdS}_{2}\otimes \mathrm{S}%
_{2}$}
\date{\today }
\author{Jorge Noronha}
\affiliation{Department of Physics, Columbia University, 538 West 120th Street, New York,
NY 10027, USA}
\affiliation{Instituto de F\'{\i}sica, Universidade de S\~{a}o Paulo, C.P. 66318,
05315-970 S\~{a}o Paulo, SP, Brazil}
\author{Gabriel S.\ Denicol}
\affiliation{Department of Physics, McGill University, 3600 University Street, Montreal,
QC, H3A 2T8, Canada}
\affiliation{Physics Department, Brookhaven National Lab, Building 510A, Upton, NY, 11973, USA}

\begin{abstract}
In this paper we obtain an analytical solution of the relativistic Boltzmann equation under the relaxation time approximation that describes the out-of-equilibrium dynamics of a radially expanding massless gas. This solution is found by mapping this expanding system in flat spacetime to a static flow in the curved spacetime $\mathrm{AdS}_{2}\otimes \mathrm{S}_{2}$. We further derive explicit analytic expressions for the momentum dependence of the single
particle distribution function as well as for the spatial dependence of its moments. We find that this dissipative system has the ability to flow as a perfect fluid even though its entropy density does not match the equilibrium form. The
non-equilibrium contribution to the entropy density is shown to be due to higher order
scalar moments (which possess no hydrodynamical interpretation) of the
Boltzmann equation that can remain out of equilibrium but do not couple to the energy-momentum tensor of the system. Thus, in this system the slowly moving hydrodynamic degrees of freedom can exhibit true perfect fluidity while being totally decoupled from the fast moving, non-hydrodynamical microscopic degrees of freedom that lead to entropy production.

\end{abstract}

\author{}
\maketitle


\section{Introduction}

The Boltzmann equation describes the underlying microscopic dynamics of
dilute classical gases \cite{cercignanibook}. It is widely employed to model
a variety of nonequilibrium phenomena in several areas of physics such as
the dynamics of the hot hadronic matter produced in the late stages of
ultrarelativistic heavy ion collisions \cite{Bass:1998ca,Petersen:2014yqa}, some aspects
of the expansion of our universe in cosmology applications \cite{bernstein},
the description of micro and nano-flows \cite{Struchtrup}, among others.

In addition to these applications, exact solutions of the relativistic
generalization of the Boltzmann equation \cite{degroot,cercignanikremer} in
the relaxation time approximation \cite{BGK,AW} have been recently employed
to improve our understanding of the domain of applicability of relativistic
dissipative fluid dynamics in the context of relativistic heavy ion
collisions \cite%
{Florkowski:2013lza,Florkowski:2013lya,Denicol:2014mca,Denicol:2014xca,Denicol:2014tha}%
. Even though less complete, the Anderson-Witting-Boltzmann (AWB) equation
and its solutions can be used to understand certain properties of solutions
of the Boltzmann equation itself, as well as its hydrodynamic limit.

Analytic solutions of the relativistic Boltzmann equation are extremely
rare (see \cite{Bazow:2015dha} for the first analytical solution in an expanding background). The same can be said even for simplified versions of the relativistic
Boltzmann equation, such as the AWB equation. Recently, an exact solution of
the AWB equation \cite{AW} was derived for a conformal system undergoing
simultaneously longitudinal and transverse expansion in \cite%
{Denicol:2014xca,Denicol:2014tha} (for an extension involving anisotropic
hydrodynamics see \cite{Nopoush:2014qba}). The remarkable agreement between
these solutions and those of relativistic dissipative fluid dynamics (under
the same symmetries) has brought great insights about the validity of the
hydrodynamic description of the evolution of the quark-gluon plasma.
However, even in this case the solutions of the AWB equation were obtained
using iterative numerical methods and it was not known how to obtain
analytic expressions for the momentum dependence of the single particle
distribution function, $f$, and the spatial dependence of its moments.

In this paper, we expand on the arguments developed in Ref.\ \cite%
{Denicol:2014xca,Denicol:2014tha} to obtain a new fully analytical solution
for the single particle distribution function of the AWB equation for
conformal kinetic systems. The key difference with respect to the exact
solutions previously derived in \cite{Denicol:2014xca,Denicol:2014tha}
involves the global symmetries imposed on the conformal system. The symmetry
assumptions \cite{Gubser:2010ze,Gubser:2010ui,Marrochio:2013wla} previously
employed in \cite{Denicol:2014xca,Denicol:2014tha} were more applicable to
the matter created in ultracentral relativistic heavy ion collisions while
in this work we broaden our focus and consider symmetries more appropriate
for conformal systems undergoing three dimensional radial expansion, such as
the early universe\footnote{An important distinction with respect to the physics of the early universe
is that here we still consider an underlying flat spacetime.}. We note that
the same set of symmetries has already been imposed to conformal fluids in 
\cite{Hatta:2014gqa,Hatta:2014gga} in order to find the first analytical
solutions of second order conformal fluid dynamics.

The possession of an analytical solution for $f$ allowed us to directly
explore here important technical aspects in kinetic theory such as the
imposition of matching conditions, the decomposition of $f$ in its moments in a nontrivial setting as well as its positivity. More importantly, this analytical solution has also
revealed a new feature of conformally invariant, radially expanding systems
described by the AWB equation: the ability to flow as a perfect fluid even
though the overall dynamics is intrinsically dissipative (e.g., the
non-equilibrium entropy component is nonzero). In fact, we show that in this
solution the energy-momentum tensor is exactly that of an ideal fluid at any
spacetime point (even though the shear viscosity coefficient is nonzero)
while the entropy density, computed directly using the full distribution
function, is different than its ideal limit. In this case, this
non-equilibrium contribution to the entropy density is due to higher order
scalar moments (which possess no hydrodynamical interpretation) of the
Boltzmann equation \cite{Denicol:2012cn} that remain out of equilibrium while the energy-momentum tensor retains its
local equilibrium form. Therefore, in the system considered here, slowly moving
hydrodynamic degrees of freedom can exhibit true perfect fluidity while
being totally decoupled from the fast moving, non-hydrodynamical microscopic
scalar degrees of freedom that lead to entropy production.

This paper is organized as follows. In the next section we briefly review how
$\mathrm{AdS}_{2}\otimes \mathrm{S}_{2}$ invariant solutions of fluid dynamics
were obtained in Refs.~\cite{Hatta:2014gqa,Hatta:2014gga}. In Sec.\ \ref{SecII} we
derive the main results of this paper and solve the Anderson-Witting-Boltzmann equation 
for a conformal system in $\mathrm{AdS}_{2}\otimes \mathrm{S}_{2}$ geometry. We show in Sec.\ \ref{SecIII}
how these solutions appear from the perspective of the method of moments. We then conclude with a summary of our results.
Throughout this paper, we use natural units $\hbar =c=k_{B}=1$.

\section{Relativistic hydrodynamics in $\mathrm{AdS}_{2}\otimes \mathrm{S}%
_{2}$}
\label{SecI}

We follow \cite{Hatta:2014gga} and consider the out-of-equilibrium dynamics
of a conformal system in $\mathrm{AdS}_{2}\otimes \mathrm{S}_{2}$ geometry.
This curved geometry is conformally equivalent to 4-dimensional Minkowski
spacetime (in spherical coordinates), 
\begin{equation}
d\hat{s}^{2}=\frac{-dt^{2}+dr^{2}+r^{2}d\Omega ^{2}}{r^{2}}=-\cosh ^{2}\rho
\,d\tau ^{2}+d\rho ^{2}+d\Omega ^{2}\,,
\end{equation}%
where $d\Omega ^{2}=d\theta ^{2}+\sin ^{2}\theta \,d\phi ^{2}$ is the usual
angular piece involving the angles $\theta \in \lbrack 0,\pi ]$ and $\phi
\in \lbrack 0,2\pi ]$ while $\tau $ and $\rho $ are global $\mathrm{AdS}_{2}$
coordinates defined using the Minkowski time, $t$, and 3-dimensional
spatial radius, $r$, in the following way \cite{Hatta:2014gga} 
\begin{equation}
\text{\ }\tan \tau =\frac{L^{2}+r^{2}-t^{2}}{2Lt},\text{ \ \ \ \ \ \ }\cosh
\rho =\frac{1}{2Lr}\sqrt{\left( L^{2}+(r+t)^{2}\right) \left(
L^{2}+(r-t)^{2}\right) }\,,
\label{definetaurho}
\end{equation}%
with $L$ being the radius of $\mathrm{AdS}_{2}$. In this curved space, quantities evolve in $\tau$ while $\rho$ plays the role of a spatial radial coordinate. In this Weyl rescaled
coordinate system the nonzero Christoffel symbols are 
\begin{equation}
\Gamma _{\theta \phi }^{\phi }=\frac{1}{\tan \theta }\,,\qquad \Gamma _{\phi
\phi }^{\theta }=-\cos \theta \,\sin \theta \,,\qquad \Gamma _{\tau \tau
}^{\rho }=\cosh \rho \,\sinh \rho \,,\qquad \Gamma _{\tau \rho }^{\tau
}=\tanh \rho \,.
\end{equation}

The energy-momentum tensor, $T^{\mu \nu }$, of a relativistic conformal
fluid is usually decomposed in terms of the time-like (normalized) local
velocity field, $u^{\mu }$, as%
\begin{equation*}
T^{\mu \nu }=\varepsilon u^{\mu }u^{\nu }+P\Delta ^{\mu \nu }+\pi ^{\mu \nu }%
\text{.}
\end{equation*}%
Above, we introduced the energy density $\varepsilon \equiv u_{\mu }u_{\nu
}T^{\mu \nu }$, the thermodynamic pressure $P\left( \varepsilon \right)
=\varepsilon /3$, and the shear stress tensor $\pi ^{\mu \nu }\equiv \Delta
_{\alpha \beta }^{\mu \nu }T^{\alpha \beta }$. We further defined the
projection operator onto the space orthogonal to $u^{\mu }$, $\Delta ^{\mu
\nu }=g^{\mu \nu }+u^{\mu }u^{\nu }$, and the double, symmetric, traceless
projection operator $\Delta _{\alpha \beta }^{\mu \nu }=\left( \Delta
_{\alpha }^{\mu }\Delta _{\beta }^{\nu }+\Delta _{\alpha }^{\nu }\Delta
_{\beta }^{\mu }\right) /2-\Delta ^{\mu \nu }\Delta _{\alpha \beta }/3$. Our
convention is to define the fluid velocity using the Landau picture, $T^{\mu
\nu }u_{\nu }=-\varepsilon u^{\mu }$, which implies that the energy diffusion
is always zero. The bulk viscous pressure of a conformal
fluid is always zero, which means that the dissipative processes involving
energy and momentum in such systems are solely governed by the shear stress
tensor.

The main equations of motion satisfied by this fluid are given by the
conservation laws of energy-momentum, which we decompose in the following
form,%
\begin{eqnarray}
u_{\nu }D_{\mu }T^{\mu \nu } &=&u^{\mu }D_{\mu }\ln T+\frac{1}{3}D_{\mu
}u^{\mu }+\frac{1}{3}\frac{\pi ^{\mu \nu }}{Ts}D_{\mu }u_{\nu }=0,
\label{eq1} \\
\Delta _{\nu }^{\lambda }D_{\mu }T^{\mu \nu } &=&u^{\mu }D_{\mu }u^{\lambda
}+\Delta ^{\lambda \mu }\partial _{\mu }\ln T+\Delta _{\nu }^{\lambda
}D_{\mu }\pi ^{\mu \nu }=0,  \label{eq2}
\end{eqnarray}%
where $D_{\mu }$ is the general relativistic covariant derivative. The
equations above are then complemented by the equations of motions for the
shear-stress tensor, $\pi ^{\mu \nu }$, which, at second order in gradients 
\cite{IS,Baier:2007ix,Denicol:2012cn}, correspond to a relaxation-type
equation%
\begin{equation}
\tau _{\pi }\Delta _{\alpha }^{\mu }\Delta _{\beta }^{\nu }u^{\lambda
}D_{\lambda }\pi ^{\alpha \beta }+\pi ^{\mu \nu }=-2\eta \sigma ^{\mu \nu }-%
\frac{4}{3}\tau _{\pi }\pi ^{\mu \nu }D_{\lambda }u^{\lambda }+\frac{10}{7}%
\tau _{\pi }\pi ^{\lambda \left\langle \mu \right. }\sigma _{\lambda
}^{\left. \nu \right\rangle }+\text{\textrm{higher-order terms} },
\label{eq3}
\end{equation}%
where\ $\eta $ is the shear viscosity and $\tau _{\pi }$ is the shear
relaxation time. For a conformal fluid, the shear viscosity must be
proportional to the entropy, $\eta \sim s$, while the shear viscosity relaxation time must
be inversely proportional to the temperature, $\tau _{\pi }\sim 1/T$. Above,
we introduced the shear tensor of the fluid, $\sigma ^{\mu \nu
}=D^{\left\langle \mu \right. }u^{\left. \nu \right\rangle }$. The brackets $%
\left\langle {}\right\rangle $ denote the transverse and traceless
projection of a tensor $A^{\left\langle \mu \nu \right\rangle }=\Delta
_{\alpha \beta }^{\mu \nu }A^{\alpha \beta }$.

The hydrodynamical solution studied in \cite{Hatta:2014gga} was constructed
using a static though non-uniform local velocity, $u_{\mu
}=(-\cosh \rho ,0,0,0)$ in $\mathrm{AdS}_{2}\otimes \mathrm{S}_{2}$ space
with coordinates $\left( \tau ,\rho ,\theta ,\phi \right) $. This implies
that the system is undergoing a certain type of spherically symmetric radial
flow in the usual Minkowski coordinates that is equivalent to the conformal soliton flow that was first introduced in \cite{Friess:2006kw} in the context of the gauge/gravity duality \cite{Maldacena:1997re} (see, e.g., \cite{Hatta:2014gga} for more details about our flow velocity in Minkowski coordinates). With this static flow configuration in $\mathrm{AdS}_{2}\otimes \mathrm{S}_{2}$, the expansion rate of the fluid vanishes, i.e., $D_\mu u^\mu =0$, and so does the shear tensor, $\sigma
^{\mu \nu }=0$. Thus, Eqs.\ (\ref{eq1}) and (\ref{eq2}) can only be
satisfied if the temperature and $\pi^{\mu\nu}$ depend solely on the spatial coordinate $\rho$, e.g., $T\left( \tau ,\rho ,\theta ,\phi \right) \rightarrow T\left( \rho
\right) $.

Moreover, note that in this space $\pi^{\mu\nu}$ is trivial: a quick look at Eq.\ \eqref{eq3} (and its generalization including terms involving higher order derivatives of the flow) reveals that in this problem $\pi^{\mu\nu}$ is identically zero. In fact, since here the flow is static and $\sigma^{\mu\nu}=0$, $D_\mu u^\mu =0$, and $\pi^{\mu\nu}=\pi^{\mu\nu}(\rho)$, in our conformal theory there are no dynamical sources available to induce a nontrivial spatial profile for the shear stress tensor, which must then vanish in all space. If nonlinear terms quadratic (or of higher-order) in $\pi^{\mu \nu }$ were present in (\ref{eq3}), nontrivial solutions of these homogeneous algebraic equations for $\pi^{\mu\nu}$ could be found \cite{Hatta:2014gqa,Hatta:2014gga} but those would necessarily assume that $\pi^{\mu \nu}$ must be nonzero for any value of $\rho$. Therefore, this nontrivial branch of solutions is not smoothly connected to the usual hydrodynamic gradient expansion for which, in this problem, the first-order Navier-Stokes contribution vanishes. In any case, this type of solutions is not going to play a role in our discussion since the nonlinear terms in $\pi^{\mu\nu}$ cannot appear in an effective hydrodynamic theory obtained from the Boltzmann equation with a linearized collision term \cite{Denicol:2010xn} such as in AWB. Therefore, one can safely set $\pi^{\mu\nu}=0$ in the following. Also, since $%
\pi _{\mu \nu }$ transforms covariantly under Weyl transformations \cite%
{Baier:2007ix}, the fact that this quantity vanishes in $\mathrm{AdS}%
_{2}\otimes \mathrm{S}_{2}$ implies that it will also vanish in Minkowski
coordinates.

In this case, the momentum
equation \eqref{eq2} leads to an equation of motion for the temperature that can be
easily solved \cite{Hatta:2014gga} 
\begin{equation}
\partial _{\rho }\ln T=-\tanh \rho \Longrightarrow T\left( \rho \right) \sim (\cosh \rho)^{-1}\,.  \label{defineT}
\end{equation}%
The interesting feature of this solution is that it corresponds to the solution of an ideal fluid. This happened without making any assumptions about the magnitude of the shear viscosity coefficient -- it simply appeared as a feature of
this highly symmetrical flow configuration. That is, even though the system
in principle has a nonzero shear viscosity coefficient, its hydrodynamic
degrees of freedom cannot dissipate since all gradients are exactly zero in $%
\mathrm{AdS}_{2}\otimes \mathrm{S}_{2}$ (note that dissipation via bulk
viscosity is forbidden due to exact conformal invariance). In Minkowski space, the temperature evolves in time as it would in a genuine, dissipationless fluid.

In the next sections we investigate the same problem of out-of-equilibrium $%
\mathrm{AdS}_{2}\otimes \mathrm{S}_{2}$ dynamics from a kinetic theory
perspective using the AWB equation. We then clarify which non-hydrodynamic
degrees of freedom of the microscopic theory are responsible for dissipation
in this case and why such degrees of freedom do not couple with the
hydrodynamic modes.

\section{Anderson-Witting-Boltzmann equation}

\label{SecII}

The \textit{on-shell} AWB equation in curved spacetime is \cite%
{Denicol:2014xca,Denicol:2014tha} 
\begin{equation}
p^{\mu }\partial _{\mu }f+\Gamma _{\mu i}^{\lambda }p_{\lambda }p^{\mu }\,%
\frac{\partial f}{\partial p_{i}}=\frac{p^{\mu }u_{\mu }}{\tau _{\mathrm{rel}%
}}\left( f-f_{\mathrm{eq}}\right) \,,  \label{AWBeq}
\end{equation}%
where the distribution function $f=f(x^{\mu },p_{i})$ is defined in a
7-dimensional phase space \cite{debbasch} in which each point is described
by seven coordinates, i.e., the $\mathrm{AdS}_{2}\otimes \mathrm{S}_{2}$
spacetime coordinates $x^{\mu }=(\tau ,\rho ,\theta ,\phi )$ and the three
spatial covariant momentum components $p_{i}=(p_{\rho },p_{\theta },p_{\phi
})$. The zeroth component of the momentum is obtained from the on-shell
condition for massless particles $p_{\mu }p^{\mu }=0$. Moreover, $f_{\mathrm{%
eq}}=\exp \left( p^{\mu }u_{\mu }/T\right) $ is the local equilibrium
distribution function for massless particles with Boltzmann statistics, $T$
is the local temperature, $u^{\mu }$ is the local velocity of the system,
and $\tau _{\mathrm{rel}}$ is the relaxation time associated with the collision operator. Conformal invariance
imposes that the relaxation time must be inversely proportional to the
temperature, $\tau _{\mathrm{rel}}=c/T$, with $c$ being a constant that is
directly related to the shear viscosity to entropy density ratio, $\eta
/s=5c $ \cite{Denicol:2010xn,Denicol:2011fa} (thus, the free streaming limit
corresponds to $c\rightarrow \infty $).

At first glance, it may appear that the AWB equation is a linear equation in 
$f$. However, we note that Eq.\ \eqref{AWBeq} must be solved simultaneously
with the equations of motion for the temperature and velocity, Eqs.\ (\ref%
{eq1}) and (\ref{eq2}). In these, one must also use the definition of the shear stress
tensor of a dilute single component gas,%
\begin{equation*}
\pi ^{\mu \nu }=T^{\left\langle \mu \nu \right\rangle }=\int \frac{d^{3}p}{%
\left( 2\pi \right) ^{3}}\frac{p^{\left\langle \mu \right. }p^{\left. \nu
\right\rangle }}{p^{\tau }\sqrt{-g}}f.
\end{equation*}%
In the end, one has a coupled set of nonlinear integro-differential
equations for $f$, $T$, and $u^{\mu }$. It is commonly very challenging to
solve these types of equations even numerically. However, as mentioned
above, exact solutions of this system of equations have been recently
obtained using iterative numerical methods \cite%
{Florkowski:2013lza,Denicol:2014xca,Denicol:2014tha}. For the type of flow
and symmetries considered in this paper, we demonstrate in the following
sections that it is possible to obtain analytic solutions of this system of
equations. We note that the collisionless limit of a system with a flow equivalent to ours in Minkowski space was previously studied in \cite{Nagy:2009eq} using very different techniques than the ones used below.

\subsection{Analytic Solution}

As mentioned in the previous section when we discussed the fluid dynamical
equations, the symmetry for the static flow imposes that $u_{\mu }=(-\cosh
\rho ,0,0,0)$. Also, for this type of static flow $f$ may depend
only on the spatial coordinates $\rho $, $\theta $, and $\phi $ (though we
shall see that $f$ does not depend on this coordinate in the end) and their
corresponding momenta.  

Since in the AWB equation the collision term is approximated to be linear in 
$f-f_{\mathrm{eq}}$, it is impossible for terms quadratic or quartic in $\pi
^{\mu \nu }$ to appear in the equation of motion for $\pi ^{\mu \nu }$ at
any order in the hydrodynamic series \cite{Denicol:2010xn,Denicol:2012cn}. Such terms can
only originate from the nonlinear terms of the collision operator and,
assuming that higher order tensorial moments \cite{Denicol:2012cn} initially
vanish, the shear stress tensor constructed using the solution $f$ of Eq.\ %
\eqref{AWBeq} must be zero. Therefore, since the bulk viscous pressure has to be
zero due to the underlying conformal invariance, the temperature that enters
the AWB equation will satisfy Eq.\ \eqref{defineT} with solution 
\begin{equation}
T\left( \rho \right) =\frac{T_{0}}{\cosh \rho }\,,
\end{equation}%
where $T_0$ is a constant. Note that this is not usually the case and in general the temperature has to be solved simultaneously with the AWB equation \cite{Denicol:2014xca,Denicol:2014tha}. The fact that the velocity profile is static and the temperature profile can be solved analytically will be extremely useful here since it will allow us
to find analytical solutions of the AWB equation for this system. These solutions for $T$ and $u^{\mu }$ serve to considerably simplify the expression for the local equilibrium distribution function and the
relaxation time, which take the following form%
\begin{eqnarray}
f_{\mathrm{eq}} &=&\exp \left[ -p^{\tau }\cosh \rho /T\left( \rho \right) %
\right] \,,  \label{important} \\
\tau _{\mathrm{rel}} &=&\frac{c}{T\left( \rho \right) }=\frac{c}{T_{0}}\cosh
\rho \,,  \label{important2}
\end{eqnarray}%
where $p^{\tau }=\sqrt{p_{\rho }^{2}+p_{\theta }^{2}+\left( p_{\phi
}^{2}/\sin ^{2}\theta \right) }/\cosh \rho $.

The AWB equation then becomes 
\begin{eqnarray}
&&p_{\rho }\partial _{\rho }f-\tanh \rho \left( p_{\rho }^{2}+p_{\theta
}^{2}+\frac{p_{\phi }^{2}}{\sin ^{2}\theta }\right) \,\frac{\partial f}{%
\partial p_{\rho }}+p_{\theta }\partial _{\theta }f  \notag \\
&+&\frac{1}{\tan \theta }\frac{p_{\phi }^{2}}{\sin ^{2}\theta }\frac{%
\partial f}{\partial p_{\theta }}=-\frac{T_{0}}{c}\frac{1}{\cosh \rho }\sqrt{%
p_{\rho }^{2}+p_{\theta }^{2}+\frac{p_{\phi }^{2}}{\sin ^{2}\theta }}%
\,\left( f-f_{\mathrm{eq}}\right) \,,  \label{Great}
\end{eqnarray}%
where we used Eq.\ (\ref{important2}). We note that $f_{\mathrm{eq}}$ itself
satisfies this equation, as is expected for a stationary solution (see also the collisionless study of \cite{Nagy:2009eq}). We also
remark that there are no terms including $\partial f/\partial p_{\phi }$,
which is consistent with spherical symmetry in these coordinates and,
thus, $f$ does not depend on $\phi $. It is then easy to see that the
general solution of this equation can be written as a sum of an equilibrium
piece and a non-equilibrium part as follows: $f(\rho ,\theta ;p_{\rho
},p_{\theta },p_{\phi })=f_{\mathrm{eq}}+f_{\mathrm{eq}}\Phi (\rho ,\theta
;p_{\rho },p_{\theta },p_{\phi })$ where the non-equilibrium piece is 
\begin{equation}
\Phi (\rho  ;p_{\rho },p_{\Omega})=\mathcal{J}\left( \frac{%
\sqrt{p_{\rho }^{2}+p_{\Omega }^{2}}\cosh \rho }{T_{0}}\right) \,\exp \left[
-\frac{T_{0}}{c}\,\frac{p_{\rho }}{|p_{\rho }|}\,\mathrm{arctan}\left( \sinh
\rho \sqrt{1+\frac{p_{\Omega }^{2}}{p_{\rho }^{2}}}\right) \right] .
\label{phisolution}
\end{equation}%
Here, $\mathcal{J}(\gamma )$ is an arbitrary function of its argument $%
\gamma $ and we have defined the short-hand notation $p_{\Omega }^{2}\equiv
p_{\theta }^{2}+\left( p_{\phi }^{2}/\sin ^{2}\theta \right) $. By taking $%
c\rightarrow \infty $, one can see that $\mathcal{J}$ is actually the
solution of this equation in the free-streaming limit, $\mathcal{J}=\Phi _{%
\mathrm{free-streaming}}$. As will be discussed in the following, the
functional form of $\mathcal{J}$ can be determined by using the matching
condition for the energy density while requiring that $f$ is
positive-definite at any point of phase space. As far as we are aware, this
is the first analytical solution of the AWB equation that describes a
radially expanding system.

\subsection{Matching condition and positivity}

In kinetic theory it is quite common to define the temperature of the system
by requiring that the energy density of the system is solely determined by
its equilibrium value,%
\begin{equation*}
\varepsilon =u_{\mu }u_{\nu }T^{\mu \nu }=\varepsilon _{\mathrm{eq}}\left(
T\right) .
\end{equation*}%
This condition implies that the following integral must always vanish 
\begin{equation}
\int \frac{\,d^{3}p}{(2\pi )^{3}}\,\frac{p^{\tau }\cosh \rho }{\,\sin \theta 
}\,f_{\mathrm{eq}}\,\Phi (\rho ,\theta ;p_{\rho },p_{\theta },p_{\phi })\,=0.
\end{equation}%
Using the analytic solution derived in the previous section, Eq.\ (\ref%
{phisolution}), it is possible to reduce this integral to a considerably
simpler form 
\begin{equation}
\int_{0}^{\infty }d\gamma \,\gamma ^{3}\mathcal{J}(\gamma )\exp \left(
-\gamma \right) =0\,.  \label{energycondition}
\end{equation}

Now, the condition (\ref{energycondition}) can be used to determine $%
\mathcal{J}(\gamma )$. For simplicity, in this work we consider a polynomial 
\textit{Ansatz} 
\begin{equation}
\mathcal{J}(\gamma )\sim a\gamma -1\,,
\end{equation}%
and one can easily find that condition (\ref{energycondition}) is met as
long as $a=1/4$. Therefore, 
\begin{equation}
\mathcal{J}(\gamma )\sim \frac{\gamma }{4}-1\,.
\end{equation}%
Note that this function is not positive-definite for $\gamma \in \lbrack
0,4] $. However, we still have the freedom to fix the overall multiplicative
constant. A mandatory physical constraint is that, in the end, the
distribution function must be a non-negative real-valued function of its
arguments. In fact, positivity can be obtained as follows. First, note that the sign of the
exponent in our solution for $\Phi $, in Eq.\ (\ref{phisolution}), is
determined by the sign of $p_{\rho }$: in the limit of $\rho \rightarrow
\infty $, the solution is bounded by $\exp \left[ -\pi T_{0}/(2c)\right] $,
when $p_{\rho }>0$, and by $\exp \left[ \pi T_{0}/(2c)\right] $, when $%
p_{\rho }<0$. To make sure that $f$ is positive-definite and, at the same
time, that $\lim_{c\rightarrow 0}f=f_{\mathrm{eq}}$ (i.e., for a vanishing
relaxation time one must recover the local equilibrium) we fix the overall multiplicative constant to be $\exp \left[
-T_{0}\pi (1+\xi )/(2c)\right] $ with $\xi >0$ and, thus, 
\begin{equation}
\mathcal{J}(\gamma )=\left( \frac{\gamma }{4}-1\right) \exp \left[ -\frac{%
\pi T_{0}}{2c}(1+\xi )\right] \,.  \label{defineJ}
\end{equation}%
In principle, other forms of $\mathcal{J}(\gamma )$ may be used in order to
achieve the same outcome, which would then generate a class of solutions of
the AWB equation. In this work, however, we limit our discussion to the form %
\eqref{defineJ} for $\mathcal{J}(\gamma )$.

It is instructive to study the dependence of $f$ on some of its arguments.
For instance, for $\rho =0$ 
\begin{equation}
\frac{f}{f_{\mathrm{eq}}}\Big|_{\rho =0}=1+\exp \left[ -\frac{\pi T_{0}}{2c}%
(\xi +1)\right] \left( \frac{1}{4T_{0}}\sqrt{p_{\rho }^{2}+p_{\theta }^{2}+%
\frac{p_{\phi }^{2}}{\sin ^{2}\theta }}-1\right)  \label{dontcare}
\end{equation}%
while for $p_{\theta }=p_{\phi }=0$ 
\begin{equation}
\frac{f}{f_{\mathrm{eq}}}\Big|_{p_{\theta },p_{\phi }=0}=1+\exp \left\{ -%
\frac{T_{0}}{2c}\left[ \pi (1+\xi )+2\,\frac{p_{\rho }\,}{|p_{\rho }|\,}%
\mathrm{Gd}(p_{\rho })\right] \right\} \left( \frac{1}{4T_{0}}|p_{\rho
}|\cosh \rho -1\right) \,,
\end{equation}%
where $\mathrm{Gd}(x)=2\tan ^{-1}\left( \exp x\right) -\pi /2$ is the
Gudermannian function. One can see that these expressions are
positive-definite and that they reduce to the equilibrium distribution in
the zero mean free path limit $c\rightarrow 0$.

\subsection{Non-equilibrium entropy}

The local entropy density is computed using the solution for $f$ as follows 
\cite{degroot} 
\begin{equation}
s=\frac{1}{(2\pi )^{3}}\int \frac{d^{3}p}{\sqrt{-g}\,p^{\tau }}u_{\mu
}p^{\mu }\,\,f\left( \ln \,f-1\right) \,.
\end{equation}%
It is easy to show that the equilibrium result is $s_{\mathrm{eq}%
}=4T^{3}(\rho )/\pi ^{2}$, which is what one would expect for an ideal
conformal gas with degeneracy factor equal to one. From the form of $f$
assumed in this paper, $f=f_{\mathrm{eq}}\left( 1+\Phi \right) $, one can
write the nonequilibrium correction to the entropy as 
\begin{equation}
\Delta s\equiv s-s_{\mathrm{eq}}=\int \frac{d^{3}p}{\sqrt{-g}\,p^{\tau }}%
u_{\mu }p^{\mu }\,f_{\mathrm{eq}}\left\{ \left( 1+\Phi \right) \ln \left[
1+\Phi \right] +\Phi \left( \ln f_{\mathrm{eq}}-1\right) \right\} \,.
\end{equation}%
The second term can be reduced to 
\begin{eqnarray}
\int \frac{d^{3}p}{\sqrt{-g}\,p^{\tau }}u_{\mu }p^{\mu }\,f_{\mathrm{eq}%
}\Phi \left( \ln f_{\mathrm{eq}}-1\right) \,\, &=&\frac{T^{3}}{4\pi ^{2}}%
\mathcal{H}(\rho )\left[ \int_{0}^{\infty }d\gamma \,\gamma ^{2}e^{-\gamma
}\left( 1+\gamma \right) \mathcal{J}(\gamma )\right] \\
&=&-\frac{T^{3}}{8\pi ^{2}}\mathcal{H}(\rho )\,\exp \left[ -\frac{\pi T_{0}}{%
2c}\left( 1+\xi \right) \right] \,,  \notag
\end{eqnarray}%
where 
\begin{equation}
\mathcal{H}(\rho )=2\int_{0}^{1}dx\,\cosh \left[ \frac{T_{0}}{c}\mathrm{%
arctan}\left( \frac{\sinh \rho }{x}\right) \right] \,  \label{defineHfinal}
\end{equation}%
is a positive-definite function. When going from the first line to the second line above, we replaced the
form of $\mathcal{J}(\gamma )$ obtained in the previous sections. The full
result is 
\begin{eqnarray}
\frac{\Delta s}{s_{eq}} &=&-\int_{0}^{\infty }\frac{d\gamma }{16}\,\gamma
^{2}\,e^{-\gamma }\int_{0}^{1}dx\,\left\{ 1+\mathcal{J}(\gamma )\,\exp \left[
-\frac{T_{0}}{c}\tan ^{-1}\left( \frac{\sinh \rho }{x}\right) \right]
\right\}  \notag \\
&\times &\ln \left\{ 1+\mathcal{J}(\gamma )\,\exp \left[ -\frac{T_{0}}{c}%
\tan ^{-1}\left( \frac{\sinh \rho }{x}\right) \right] \right\}  \notag \\
&&-\int_{0}^{\infty }\frac{d\gamma }{16}\,\gamma ^{2}\,e^{-\gamma
}\int_{0}^{1}dx\,\left\{ 1+\mathcal{J}(\gamma )\,\exp \left[ \frac{T_{0}}{c}%
\tan ^{-1}\left( \frac{\sinh \rho }{x}\right) \right] \right\}  \notag \\
&\times &\ln \left\{ 1+\mathcal{J}(\gamma )\,\exp \left[ \frac{T_{0}}{c}\tan
^{-1}\left( \frac{\sinh \rho }{x}\right) \right] \right\} -\frac{\mathcal{H}%
(\rho )}{32}\,\exp \left[ -\frac{\pi T_{0}}{2c}(1+\xi )\right] .
\label{entropyprod}
\end{eqnarray}

\begin{figure}[t]
\includegraphics[width=0.6\linewidth]{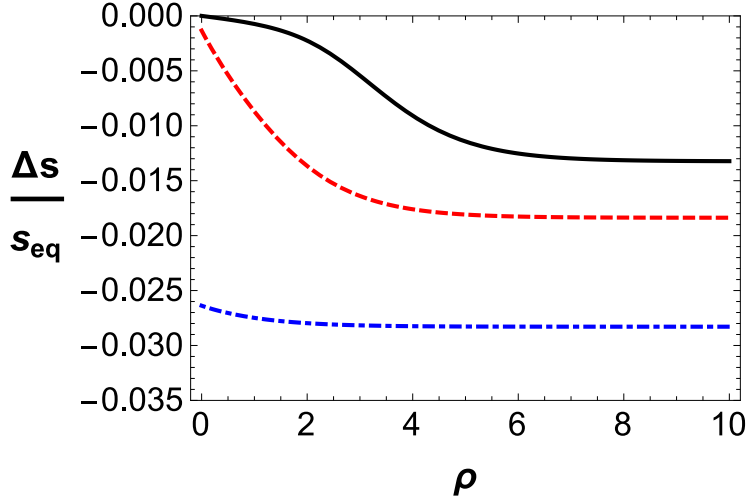}
\caption{(Color online) Relative entropy production $\Delta s/s_{eq}$ in
Eq.\ \eqref{entropyprod} for different values of $T_0/c$ (with fixed $%
\protect\xi=0.01$). The solid black line was computed using $T_0/c=10$, the
dashed red curve is for $T_0/c=1$, while the dotted-dashed blue curve is for 
$T_0/c=0.1$.}
\label{fig:1}
\end{figure}

It is easy to see that $\Delta s(\rho )$ is even in $\rho$ and that $\Delta
s(\rho )<0$, as expected on physical grounds. We show in Fig.\ \ref{fig:1} a
plot of $\Delta s/s_{eq}$ as a function of $\rho$ for different values of $%
T_0/c$. For small values of $T_0/c$ one can see that the full entropy
density becomes different than the equilibrium one (though by a small
amount) and that this effect becomes more pronounced for large values of $%
\rho$, where it reaches a stationary value that depends on the parameters $c$
and $T_0$.

Note that $\Delta s/s_{eq}$ in Eq.\ \eqref{entropyprod} does not change under Weyl transformations and, thus, one can find its value in flat spacetime via simple substitution $\frac{\Delta s}{s_{eq}}(\rho) = \frac{\Delta s}{s_{eq}}(\rho(t,r))$. 
Also, since large values of $\rho$ correspond to large values of $t$ for fixed $r$ (see Eq.\ \eqref{definetaurho}), this quantity approaches a constant at large times in flat spacetime. From the result shown in Fig.\ \ref{fig:1}, one can see that the spatial integral of $\Delta s$ in flat spacetime 
\begin{equation}
 \frac{128 L^3 T_0^3}{\pi} \int_0^\infty dr\,\frac{r^2}{\left[ L^2 +(r+t)^2   \right]^{3/2}\left[ L^2 +(r-t)^2   \right]^{3/2}}\,\frac{\Delta s}{s_{eq}}\Big|_{\rho=\rho(t,r)}
\end{equation}
goes to zero when $t \to \infty$, which indicates that the entropy approaches its equilibrium value as time increases. The equation above was obtained using that the equilibrium entropy density in flat spacetime is $s_{eq}(t,r) = 4 T^3(t,r)/(\pi^2 r^3)$.

\section{Comparison to the method of moments}

\label{SecIII}

In order to better understand some features of the solution derived in the
previous sections it is convenient to expand $\Phi =\left( f-f_{\mathrm{eq}%
}\right) /f_{\mathrm{eq}}$ in terms of its moments, using irreducible
tensors and a complete basis of polynomials \cite{Denicol:2012cn}. The
irreducible tensors, $1$, $k^{\left\langle \mu \right\rangle }$, $%
k^{\left\langle \mu \right. }k^{\left. \nu \right\rangle }$, $%
k^{\left\langle \mu \right. }k^{\nu }k^{\left. \lambda \right\rangle }$, $%
\cdots $, are used to expand the angular part of the single-particle
distribution function. They form a complete and orthogonal set, analogously
to the spherical harmonics \cite{degroot}, and are defined as $%
k^{\left\langle \mu _{1}\right. }...k^{\left. \mu _{m}\right\rangle }\equiv
\Delta _{\nu _{1}...\nu _{m}}^{\mu _{1}...\mu _{m}}k^{\nu _{1}}...k^{\nu
_{m}}$, where the transverse, symmetric, and traceless projectors $\Delta
_{\nu _{1}...\nu _{m}}^{\mu _{1}...\mu _{m}}$ are defined in \cite{degroot}. Our solution in Eq.\ \eqref{phisolution} is anisotropic in momentum space and hence it possesses both scalar and higher rank moments. For the sake of illustration, in this section we focus on the scalar moments of our solution. The scalar part of the distribution function is expanded using a set of
orthogonal polynomials, $P_{\mathbf{k}n}^{(\ell
)}=\sum_{r=0}^{n}a_{nr}^{(\ell )}\left( -u_{\mu }k^{\mu }\right) ^{r}$,
where the coefficients $a_{nr}^{(\ell )}$ were calculated so that%
\begin{equation}
\frac{N^{\ell }}{\left( 2\ell +1\right) !!}\int \frac{dK}{\sqrt{-g}}\left(
u_{\mu }k^{\mu }\right) ^{2\ell }P_{\mathbf{k}n}^{(\ell )}P_{\mathbf{k}%
m}^{(\ell )}=\delta _{nm},
\end{equation}%
using the Gram-Schmidt orthogonalization method as demonstrated in \cite%
{Denicol:2012cn}. Here, we defined $dK=d^{3}k/\left[ \left( 2\pi \right)
^{3}k^{\tau}\right] $ and $N_{\ell }=(-1)^{\ell }/I_{2\ell ,\ell }$ where, for
a nongenerate massless gas of particles, 
\begin{equation*}
I_{nq}=\frac{\left( n+1\right) !}{\left( 2q+1\right) !!}\frac{T^{n+2}}{2\pi
^{2}}.
\end{equation*}%
The irreducible tensors also satisfy orthogonality conditions,%
\begin{equation}
\int \frac{dK}{\sqrt{-g}}F_{\mathbf{k}}\,k^{\left\langle \mu _{1}\right.
}\cdots k^{\left. \mu _{m}\right\rangle }\,k_{\left\langle \nu _{1}\right.
}\cdots k_{\left. \nu _{n}\right\rangle }=\frac{m!\,\delta _{mn}}{\left(
2m+1\right) !!}\,\Delta _{\nu _{1}\cdots \nu _{m}}^{\mu _{1}\cdots \mu
_{m}}\int \frac{dK}{\sqrt{-g}}\frac{N^{\ell }}{\left( 2\ell +1\right) !!}\,F_{\mathbf{k}}\left(
u_{\mu }k^{\mu }\right) ^{2m},  \label{orthogonality1}
\end{equation}%
where $F_{\mathbf{k}}$ is an arbitrary function of $u_{\mu }k^{\mu }$.

Using this basis, the moment expansion of $\Phi $ is%
\begin{equation*}
\Phi =\sum_{\ell =0}^{\infty }\sum_{n=0}^{\infty }\mathcal{P}_{\mathbf{k}%
n}^{(\ell )}\Theta _{n}^{\mu _{1}\cdots \mu _{\ell }}k_{\left\langle \mu
_{1}\right. }\cdots k_{\left. \mu _{\ell }\right\rangle },
\end{equation*}%
where the moments can be obtained using the orthogonality
relations satisfied by the basis elements and are given by 
\begin{equation}
\Theta _{n}^{\mu _{1}\cdots \mu _{\ell }}=\int \frac{d^{3}k}{(2\pi
)^{3}\,\sqrt{-g}}\,\frac{(-k\cdot u)^{n}}{k^\tau}k^{\left\langle \mu _{1}\right. }\,\cdots
k^{\left. \mu _{\ell }\right\rangle }f_{\mathrm{eq}}\Phi \,.  \label{moment}
\end{equation}%
For the sake of convenience, we defined above%
\begin{equation*}
\mathcal{P}_{\mathbf{k}n}^{(\ell )}\equiv \frac{N_{\ell }}{\ell !}%
\sum_{m=n}^{\infty }a_{mn}^{(\ell )}P_{\mathbf{k}}^{\left( n\ell \right) }.
\end{equation*}

We note that the scalar moments can also be
calculated analytically, by replacing Eq.\ (\ref{phisolution}) into Eq.\ (\ref%
{moment}). The solution is 
\begin{equation}
\Theta _{n}=\frac{n-2}{16\pi ^{2}}\,T^{n+2}(\rho )\,\Gamma (n+2)\,\mathcal{H}%
(\rho )\,\exp \left[ -\frac{\pi T_{0}}{2c}\left( 1+\xi \right) \right] \,,
\label{resultscalarmoments}
\end{equation}%
where $\Gamma (n)$ is the Gamma function. Note that this quantity vanishes for $n=2$, as expected, from the energy matching condition and that $\Theta_0/T^2=\Theta_1/T^3 <0$ while $\Theta_n >0$ for $n>2$. For the sake of completeness, in Fig.\ 2 we plot $\Theta_3/T^5$ for different values of $T_0/c$ with $\xi =0.01$. 

\begin{figure}[t]
\includegraphics[width=0.6\linewidth]{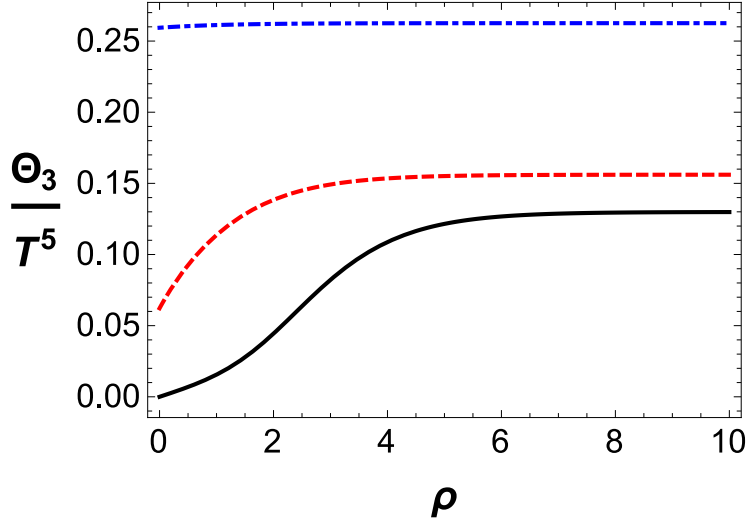}
\caption{(Color online) Normalized scalar moment $\Theta_3/T^{5}$ for different values of $T_0/c$ (with fixed $\protect\xi=0.01$). The solid black line was computed using $T_0/c=10$, the
dashed red curve is for $T_0/c=1$, while the dotted-dashed blue curve is for 
$T_0/c=0.1$.}
\label{fig:2}
\end{figure}

The actual moment expansion of $\Phi $ then becomes%
\begin{equation*}
\Phi =\sum_{n=0}^{\infty }\mathcal{P}_{\mathbf{k}n}^{(0)}\Theta _{n}\text{ }.
\end{equation*}%
Truncating this expression at $n=2$ (note that the matching condition fixes $%
\Theta _{2}=0$), we obtain something analogous to the 14-moment
approximation \cite{IS}, 
\begin{equation*}
\Phi =\mathcal{P}_{\mathbf{k}0}^{(0)}\Theta _{0}+\mathcal{P}_{\mathbf{k}%
1}^{(0)}\Theta _{1}.
\end{equation*}%
For a gas of nondegenerate massless particles, it is easy to show that%
\begin{equation*}
\mathcal{P}_{\mathbf{k}0}^{(0)}=\frac{2\pi ^{2}}{T^{2}}\left( 3+\frac{1}{T}%
u_{\mu }k^{\mu }\right) \text{, }\mathcal{P}_{\mathbf{k}1}^{(0)}=-\frac{2\pi
^{2}}{T^{3}}\left( 1+\frac{1}{2T}u_{\mu }k^{\mu }\right) \text{ }.
\end{equation*}%
where we used that%
\begin{equation*}
a_{00}^{(0)}=1\text{, }\left[ a_{11}^{(0)}\right] ^{2}=\frac{1}{2T^{4}}\text{%
, \ }\frac{a_{10}^{(0)}}{a_{11}^{(0)}}=-2T\text{.}
\end{equation*}

In this truncation scheme, the distribution function is then approximated to
be%
\begin{equation*}
\Phi =-\frac{1}{2}\,\mathcal{H}(\rho )\,\exp \left[ -\frac{\pi T_{0}}{2c}%
\left( 1+\xi \right) \right] \left( 1-\frac{1}{4T_{0}}\sqrt{p_{\rho
}^{2}+p_{\theta }^{2}+\left( p_{\phi }^{2}/\sin ^{2}\theta \right) }%
\,\right) \text{ }.
\end{equation*}%
For $\rho =0$, this is exactly the same as our analytical solution, see Eq.\ (%
\ref{dontcare}). This shows that a finite number of scalar moments are able
to provide a reasonable description of this system at least when $\rho=0$.

\section{Conclusions}

\label{SecIV}

In this paper we derived the first analytical solution of the
Anderson-Witting-Boltzmann equation for a radially expanding system (known as conformal soliton flow) of massless particles. We further demonstrated how the matching conditions, commonly used to define temperature in kinetic theory,
restrict the form of the solution of the single particle distribution
function. 

The solution we found has some very interesting features. In this
system the slowly moving hydrodynamic degrees of freedom do not see
dissipation, e.g., sound waves propagate without any distortion from viscosity.
However, faster degrees of freedom are still present and they produce a
finite amount of entropy. This may be the first example of a kinetic system that
does not have a viscous hydrodynamic behavior: between its ideal fluid and
free-streaming limits, there is no region in space and time where a viscous fluid dynamical
description is valid. 

This conclusion regarding the perfect fluidity of the conformal soliton flow, studied here at weak coupling in the context of kinetic theory, was also found in the case of an infinitely coupled $\mathcal{N}=4$ Supersymmetric Yang-Mills plasma \cite{Friess:2006kw}. In fact, even though this strongly-coupled system has nonzero shear viscosity $\eta/s=1/(4\pi)$ \cite{Kovtun:2004de}, the underlying symmetries of the flow together with conformal invariance impose that the energy-momentum tensor of the system retains its perfect fluid form. This shows that the exact cancellation of shear viscous effects in the energy-momentum tensor discussed here also happens in strongly coupled systems. 

For any finite value of $c$ in the relaxation time \eqref{important2}, our solution for the distribution function does not return to local thermal equilibrium even at sufficiently large times. In fact, one can see from \eqref{definetaurho} that large times (for fixed radius $r$) correspond to large $\rho$'s and, in this case, the non-equilibrium contribution given by \eqref{phisolution} and \eqref{defineJ} does not vanish if $c\neq 0$. Thus, in our system the effects of the expansion  overcome the collision term and the distribution function does not relax to its equilibrium form. We note that a similar conclusion was found for a different type of rapidly expanding gas in \cite{Bazow:2015dha}, which went beyond the relaxation time approximation and took into account the full nonlinearities of the collision term of the relativistic Boltzmann equation.

The essential approximations made here to find this novel many-body effect
were: relaxation time approximation, conformal dynamics, and spherical symmetry
(implemented via the $\mathrm{AdS}_{2}\otimes \mathrm{S}_{2}$ construction).
The effects discussed in this paper may appear when describing a perfectly radially symmetric and homogeneous droplet of quark-gluon plasma, at very high temperatures and
vanishing chemical potentials, expanding in vacuum. In this limit, QCD is approximately conformal and the flow configuration should resemble the one discussed
in this paper. 

\section*{Acknowledgements}

The authors thank Y.~Hatta, B.~Xiao, and M.~Martinez for collaboration in the early stage of this
work. G.~S.~Denicol is currently supported under DOE Contract No. DE-SC0012704 and acknowledges previous support of a Banting fellowship provided by the Natural Sciences and Engineering Research Council of Canada.
J.~N.\ thanks Columbia University's Physics Department for the hospitality
and Conselho Nacional de Desenvolvimento Cient\'{\i}fico e Tecnol\'{o}gico
(CNPq) and Funda\c{c}\~{a}o de Amparo \`{a} Pesquisa do Estado de S\~{a}o
Paulo (FAPESP) for financial support.



\end{document}